\documentclass[prx,floatfix,twocolumn,showkeys,showpacs]{revtex4-2}
\usepackage{amsmath}
\usepackage{graphicx}
\usepackage{amsfonts}
\usepackage{amssymb}
\usepackage{bm}
\usepackage{epsfig,float,afterpage}
\usepackage[colorlinks,linkcolor=blue,citecolor=blue,urlcolor=blue]{hyperref}
\usepackage[usenames,dvipsnames]{color}
\usepackage{lipsum}
\def\beq{\begin{equation}}
\def\eeq{\end{equation}}
\def\bea{\begin{eqnarray}}
\def\eea{\end{eqnarray}}
\def\nn{\nonumber}
\def\ba{\begin{array}}
\def\ea{\end{array}}

\newcommand{\dg}{\dagger}
\def\one{1\hskip -1mm{\rm l}}
\newlength{\sizeonefig}
\newlength{\sizetwofig}
\begin{document}

\author{Averi Banerjee} 
\affiliation{Department of Basic Science and Humanities, Techno International New Town, Kolkata-700156, India}
\author{Syeda Rafisa Rahaman}
\affiliation{Integrated Science Education \& Research Centre, Visva-Bharati University, Santiniketan-731235, India}
\author{Nilanjan Bondyopadhaya} 
\email{nilanjan.iserc@visva-bharati.ac.in}
\affiliation{Integrated Science Education \& Research Centre, Visva-Bharati University, Santiniketan-731235, India}
\title{ Transport in extended Kitaev chain with time reversal symmetry breaking and long-range interaction}

\begin{abstract}

We consider a junction consisting of an extended one-dimensional Kitaev chain which incorporates both time-reversal symmetry (TRS) breaking and long-range interaction, sandwiched between two metallic leads from two sides. In this hybrid device, we study electrical transport under voltage bias for varying strength of the TRS breaking phase. We compare the transport characteristics of long-range type Kitaev chain with that of the short-range Kitaev chain as the strength of the TRS breaking phase varies. We find that the TRS breaking modifies the density of states and localisation/delocalisation property of the eigenstates which in turn affect the transport characteristics. Moreover, we find that the impact of the TRS breaking is not identical for the long-range Kitaev chain and its short-range counterpart. Therefore, noticeable differences in the transport properties can be observed due to the interplay between the TRS breaking and the range of interaction.

\end{abstract}

\vspace{0.5cm}
%\date{\today}
\keywords{ Time reversal symmetry breaking, Long-range interaction, Kitaev chain, Transport}

\maketitle

%\section{Abstract}

\section{Introduction}
In condensed matter physics, topological phases of matter have drawn considerable interest over the past decades owing to their non-trivial exotic properties like non-locality, quantised response to external field which is immune from low-scale perturbation \cite{Qi_RMP,Hasan_2010,Bernevig2013TopologicalIA,Ramos_2022}. Topological superconductors are one such exotic system which hosts low-energetic topological states. These topological superconductors can be classified into different symmetry classes according to different symmetry properties like particle exchange symmetry, spin-rotation symmetry, particle-hole symmetry and time-reversal symmetry (TRS) \cite{Altland_PRB_1997,Fidkowski_PRB2011,Degottardi_PRB2013}. One-dimensional topological superconductors obeying time-reversal symmetry belong to class BDI which is characterised by $\mathbb{Z}$ topological invariant. On the other hand, class D comprises those one-dimensional topological superconductors which break time-reversal symmetry. Time reversal breaking term in the 1D Hamiltonian reduces the topological invariant to $\mathbb{Z}_2$ class \cite{Degottardi_PRB2013,Kotetes_2013}.

In general, time-reversal symmetry breaking in superconducting systems has also attracted significant attention in recent years. Extensive theoretical and experimental research are being carried out to understand the effect of TRS breaking in superconductors which in most cases happen due to the external magnetic fields. However, there are also several superconducting materials which spontaneously break time-reversal symmetry below critical superconducting transition temperature $T_c$ \cite{Kallin_2016,Ghosh_2021,condmat4020047,Andersen2024}. Experimentally, it is possible to directly detect the presence of spontaneous magnetic field, as well as TRS breaking in the superconducting state through $\mu SR$, the optical Kerr effect and SQUID magnetometry \cite{Ghosh_2021}.

Measurement processes based on quantum transport phenomena also offer useful tools to investigate a nonequilibrium mesoscopic system in greater detail. Hence, investigating the effect of TRS breaking on transport measurement may reveal some interesting features of the TRS breaking phase of the superconductor. 

In the paper, we choose the extended Kitaev chain which is a simple one-dimensional model of topological superconductor with time reversal symmetry breaking phase to study the effect of TRS breaking on the transport through the same. Since the extended Kitaev chain incorporates both long-range interactions and time reversal breaking phases \cite{AleccePRB2017}, it is an ideal playground to investigate both individual and combined effects of long-range interaction and time-reversal symmetry-breaking on transport characteristics, namely conductance and currents. The effect of TRS breaking on quantum transport has already been studied for different models like continuous-time quantum walk model \cite{Wong2015QuantumWS,zoltain_2013}; however, its effect on transport through the extended Kitaev chain is yet to be investigated systematically. 

We use nonequilibrium Green's function techniques based on the quantum Langevin equation to calculate electrical currents and differential electrical conductance for different TRS breaking phases \cite{DharSenPRB2006,RoyPRB2012}. Further tuning the strength of the TRS breaking phase term, we look into the effect of the TRS breaking on transport. The model we discuss here contains the TRS breaking phase in the hopping terms. This extended Kitaev chain also includes long-range interactions in both hopping and superconducting pairing terms. It has already been reported that the long-range interactions significantly modify the phase diagram, density of states and also give rise to massive subgap topological states of the long-range Kitaev chain, which in turn affect the transport characteristics \cite{VodolaPRL2014,Viyuela_PRB2016,AleccePRB2017,VodolaNJP2016,
Patrick_PRL2017,Giuliano_PRB2018,Maity_2020,Nehra_2020,BondyopadhayaJPC2024,cinnirella2024}. 
We study the effect of the TRS breaking on both long-range Kitaev (LRK) and short-range Kitaev (SRK) chains. $I$-$V$ characteristics of the extended Kitaev chain vary as the strength of the TRS breaking phase changes. Especially for a particular value of the TRS breaking phase that corresponds to its maximum values, LRK and SRK chains show distinguishable behaviour. 

%can be regulated between TRS breaking and TRS preserving 
%
%with TRS breaking phase in hopping term and long-range 

We  organise the paper as follows. A detailed discussion of the model Hamiltonian and a brief theoretical formalism is provided in the next section \ref{sec:Hamiltonian}. In Sec. \ref{sec:currents}, we discuss the analytical expressions and other details relevant for calculating electrical currents and differential conductivity. Sec. \ref{sec:results} is the result section in which we plot the results of the numerical simulations and analyse the results. We discuss the effect of the long-range interactions and time-reversal breaking phase differential electrical conductance, localization and electrical currents. Finally, we conclude in Sec. \ref{sec:con}.

\section{Model Hamiltonian and Formalism}
\label{sec:Hamiltonian}
Following Ref. \cite{AleccePRB2017}, we consider an extended version of the one-dimensional Kitaev chain which includes long-range interactions in both hoping and $p$-wave superconducting pairing terms in addition to the time-reversal symmetry (TRS) breaking phase in hopping. Hence, the Hamiltonian of the extended Kitaev chain in open configuration reads
\bea
\mathcal{H}_G&=&-\hbar \left[\sum_{k=1}^{N} \sum_{l=1}^{N-k} \left( \gamma_l  e^{i \phi_l} \, c^{\dg}_{k}c_{k+l} + \Delta_l\, c^{\dg}_{k}c^{\dg}_{k+l} + \text{H.c.} \right) \right. \nn \\
&& \left. ~ -{\epsilon} \sum_{k=1}^{N}(2c^{\dg}_{k}c_{k}-1) \right]\, ,
\label{HLRK}
\eea

where $c_l$ is the fermionic annihilation operator at $l$-th site, $N$ is the number of lattice sites of the chain and $\epsilon$ is the on-site energy. $e^{i \phi_l}$ is the TRS breaking phase in hopping which breaks time-reversal symmetry for all values of $\phi_l$, except $\phi_l=0, n\pi \, ( n \in \mathbb{Z}), \forall l$.

In this model, both hopping and pairing coupling interact through long-range interaction and show power law decay behaviour as the distance between two sites changes. We assume that the extended hopping and pairing coupling take the following forms,
\beq
\gamma_l = \gamma_0  l^{-\alpha}, ~\Delta_l = \Delta_0 l^{-\beta} \nn
\eeq
while these couple $k$-th with $(k+l)$-th sites. The parameters $\alpha \, (\alpha >0) $ and $\beta \, (\beta >0) $ control the power-law decay of the hopping and pairing terms, respectively. Ideally, long-range Kitaev (LRK) chain reduces to short-range Kitaev (SRK) chain in the limit: $\alpha \rightarrow \infty$ and $\beta \rightarrow \infty$. However, for all practical purposes, we choose $\alpha =10$ and $\beta=10$, which reduces the LRK chain to the SRK chain.
Conversely, in $(\alpha <1,\beta <1)$ region of parameter space, interactions maintain their purely long-range characteristics. It is also known that the long-range interaction affects the energy spectrum in two different ways, namely (i) it modifies the density of states \cite{BondyopadhayaJPC2024}  and (ii) it generates finite mass for topological edge modes \cite{VodolaNJP2016}. In this article, we choose $\alpha =\beta = 0.5$ while studying the LRK chain.

1D Kitaev chain with TRS protects the number of Majorana modes, thus it falls in BDI class, whereas the 1D Kitaev chain with broken TRS only preserves the {\it parity} of the number of Majorana modes and let it be classified into D class \cite{Fidkowski_PRB2011,Degottardi_PRB2013}. To study the effect of time-reversal symmetry breaking on transport phenomenon, we consider site-dependent TRS breaking phase: $\phi_l=\phi_0 l$, with $l=1,\dots, N$ \cite{AleccePRB2017}. 

%$\phi_r$: (i) constant phase : $\phi_r=\phi_0$, $\forall r =1, \dots, N$, and (ii) site dependent phase: $\phi_r=\phi_0 r$, with $r=1,\dots, N$ \cite{PhysRevB.95.195160}. 

We use Langevin equations and Green’s function method (LEGF) \cite{DharPRB2003,DharSenPRB2006,Roy_Dhar_PRB2007, RoyPRB2012,BhatDharPRB2020,BondyopadhayaJSP2022} to study differential conductance and electrical currents through the extended Kitaev chain $\mathcal{H}_G$ (\ref{HLRK}),  when the system is in nonequilibrium steady-state. 
It is well-known that the LEGF method can be employed to study nonequilibrium transport in open system configuration of the  middle wire (system) which is connected to two equilibrium reservoirs at two ends. In this case, the extended Kitaev (TS) chain is coupled with two metallic (N) reservoirs at two ends of the middle chain as shown in the Fig. \ref{cartoon}. We assume that the reservoirs are at equilibrium for time $t_-$, and their equilibrium distributions are defined by their respective chemical potentials and temperatures. The left reservoir is kept at chemical potential $\mu_L$, and temperature $T_L$ while the  right reservoir is kept at chemical potential $\mu_R$, and temperature $T_R$. In tight-binding approximation, the Hamiltonian of the  semi-infinite metallic reservoir reads
\bea
 \mathcal{H}_M^p &=& - \gamma_p \hbar \sum_{r=1}^{\infty} \left(c^{p \dg}_\alpha \, c^p_{\alpha+1}+c^{p \dg}_{\alpha+1} \, c^p_{\alpha} \right) \, ,
\label{HBath}
\eea
where $\gamma_p$ represents the hopping strength of the $p$-th metallic reservoirs and $p=L\, (R)$ represents the left (right) reservoir. $c^{p}_r$ represents an electron annihilation operator on the $r$-th site of the $p$-th reservoir. The initial equilibrium correlations for these semi-infinite metallic reservoirs are given by
\\
\begin{equation}
\langle c^{p \dg}_\alpha (t_0)c^p_{\beta} (t_0) \rangle =\sum_k \psi^{p \,*}_k(\alpha)\psi^{p }_k(\beta) f( \epsilon_k^p,\mu_p,T_p), 
\label{corrl}
\end{equation}
\\
where $\psi^{p }_k$ is the $k$-th eigenfunction of $p$-th metallic reservoir satisfying:  $\sum_{\beta} (\mathcal{H}_M^p)_{\alpha \beta}\, \psi^{p }_k(\beta)=\epsilon_k^p \,\psi^{p }_k(\alpha)$.  Clearly, $f(\omega,\mu,T)=(\exp\{(\omega -\mu)/k_B T \}+1)^{-1}$ denotes the Fermi function \cite{DharSenPRB2006} .

We connect the metallic reservoirs with the middle wire at a later time $t_0\,(t_0>t_-)$ through the tunnel junctions $\Gamma_p$ given by
\bea
\Gamma_p &=& - \gamma_p ' \hbar \, \left( c^{p \dg}_1 \, c_{l_p}+ c^\dg_{l_p} c^p_1 \right) \, ,
\label{HTun}   
\eea
\\
where $l_L=1$ and $l_R=N$ for a middle wire with $N$ sites and  $p=L\, (R)$ represents the left (right) reservoir. Here $\gamma_p '$ controls the strength of the tunnel coupling between the 
$p$-th reservoir and the middle chain.
Finally, by taking the limit $t_0 \rightarrow -\infty$, we allow the system to reach a nonequilibrium steady-state (NESS). In NESS, the Heisenberg equations are written for the whole system, i.e. reservoir-wire-reservoir. The reservoir's degrees of freedom are eliminated by using reservoir Green's functions. The equations of motion of the wire alone take the form of quantum Langevin equations in which the effect of the reservoirs appears as noise and dissipative terms. Finally, one can solve the quantum Langevin equations at $t\, (t>t_0)$ by Fourier transformations to obtain the steady-state properties of the wire. An overview of the quantum Langevin equations and Green’s function method (LEGF) is added in Appendix~\ref{App1}

%
%
%\end{section}

\section{Differential Electrical Conductance and electrical currents }
\label{sec:currents}

\begin{figure}[htb]
\centering{\includegraphics[width=1\linewidth]{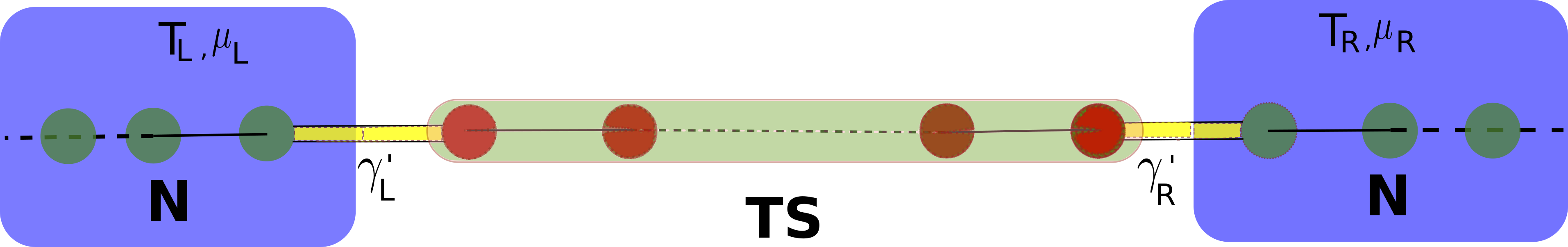}}
%\centering{\includegraphics[width=1\textwidth]{Elec_cur_vs_V_epsi_0_gamma_05.pdf}}
\caption{Sketch of N-TS-N junction}
\label{cartoon}
\end{figure}
In a N-TS-N junction, two metallic reservoirs are connected to the middle superconducting wire through two metallic tunnel couplings (see Fig. \ref{cartoon}). Conservation of electric charges are maintained throughout the metallic tunnel coupling. Hence, using the continuity equations, we can obtain the expression for electrical currents for both left and right junctions as follows:
\bea
J_{\rm L}^e &=&i e \, \gamma'_{L} \langle (c_{1}^{\dg} (t)c_{1}^L(t)-c_{1}^{L \dg}(t) c_{1}(t))\rangle \nn \\
& = & -2 e \gamma'_{L} \text{Im} [\langle c_{1}^{\dg} (t)c_{1}^L (t) \rangle ] \, , \nn\\
J_{\rm R}^e &=&i e \, \gamma'_{R} \langle (c_{1}^{R\, \dg} (t)c_{N}(t)-c_{N}^{ \dg}(t) c_{1}^{R}(t))\rangle \nn \\
&=& -2 e \gamma'_{R} \text{Im} [\langle c_{1}^{ R \, \dg} (t)c_{N} (t) \rangle ] \, , \label{elec_c}
\eea
\\
where $e$ is the magnitude of the electronic charge. The expectation $\langle .. \rangle$ denotes averaging over the initial density matrix of the reservoirs. Here, $J_L^e$ ($J_R^e$) represents the electrical current from the left reservoir to the wire (wire to the right reservoir). 

Differential electrical conductance (DEC) is a useful observable for measuring the local density of states. To calculate DEC, one first needs to determine the electrical current across the junction under voltage biasing. Under symmetric biased condition i.e. $\mu_L=-\mu_R=eV/2$ ($\mu_L-\mu_R=eV$), one can obtain the  expression for DEC at the left (right) junction by calculating $2\left[\frac{dJ^e_p}{dV}\right]$ for $p=L \,(R)$. We follow the procedure described in Ref. \cite{RoyPRB2012,BondyopadhayaJSP2022} to calculate DEC from the junction current. Denoting $\mu_L=-\mu_R=\mu$, DEC in a symmetrically biased N-TS-N junction can also be expressed as 
$e\left[\frac{dJ^e_p}{d\mu}\right]$.

Voltage biasing between the two metallic reservoirs drives the middle wire into a nonequilibrium steady-state in the absence of bound-states whose energies lie beyond the bandwidth of the reservoirs \cite{DharSenPRB2006}. In our simulations, we choose the parameters of the wire and reservoirs in a way that does not cause the bound-state related problem to arise here. It can be noted that under symmetric biasing $(\mu_L=-\mu_R= \mu)$, equality between left and right currents holds, i.e. $J^e_L=J^e_R=J^e$. We explore the effects of TRS breaking on quantum transport through the N-TS-N junctions in the presence of both long and short range type interactions in Kitaev chain. In this article, we explore two well-known transport properties, namely (i) differential electrical conductances at zero bias and (ii) electrical current ($J^e$) as a function of biasing voltage ($eV$). In all calculations in the subsequent sections, we choose hopping strengths are the same for two reservoirs, i.e. $\gamma_L=\gamma_R$, and these are connected with the middle wire by identical contacts ($\gamma_L '=\gamma_R '$).  For simplicity, we assume the TRS breaking phase has no effect on the superconducting pairing terms ($\Delta_0$), and the strength of the voltage bias is moderate enough not to cause any non-uniformity in the $\Delta_0$ throughout the superconducting wire. The whole procedure is followed at zero temperature or near zero temperature so that the $\Delta_0$ remains almost constant and the superconducting gap remains open. In all our simulations, we choose $e=1$ and $\hbar=1$. For junction current simulation, we fix $T_L=T_R=0.02$; however, we calculate DEC at zero temperature. 

%neglect any influence of voltage bias on the superconducting pairing term $\Delta_0$ because the essential qualitative features of the steady-state transport in such a device do not depend on the exact value of the $\Delta_0$ \cite{Lobos_NJP2015}.

\begin{section}{Result and Discussions}
\label{sec:results}

\begin{subsection}{Differential Electrical Conductance}
It is well known that zero-temperature DEC at junctions is an important observable to identify the emergence of Majorana bound state (MBS) in various junctions of N-TS  and N-TS-N devices. It is proposed that the existence of MBS can be identified by the appearance of quantised peak of height $2 e^2/\hbar$ within the superconducting pairing gap in the zero-temperature DEC when the TS wire is in the perfect topological phase \cite{Sau_PRB2010,Alicea_2012,Leijnse_2012,RoyPRB2012,Sau2020CountingOM,
Zhu_Science,Dvir_2023}.

We investigate the behaviour of zero-temperature DEC with respect to the variation of TRS breaking phase $\phi_0$. This variation of $\phi_0$ directly affects the DEC in both SRK and LRK chains. Let us first begin with the SRK chain. In Fig. \ref{SRK_dIdV_V_0_025_01}, we compare the zero-temperature DEC for $\phi_0=0,\pi/4,\pi/3,\pi/2$ in two different values of tunnel coupling i.e. $\gamma'_p=0.1,0.25$. We also draw Fig. \ref{SRK_dos} to depict the variation of density of states (DOS) with $\phi_0$. Fig. \ref{SRK_dIdV_V_0_025_01}(a) for $\phi_0=0$ shows quantised zero-bias peak (ZBP) in DEC ($2$ in the unit of $\frac{e^2}{\hbar}$) which is the key characteristic of MBS. A larger value of $\gamma'_p$ i.e. stronger tunnel coupling, broadens the peak width through quasi-particle poisoning \cite{Alicea_2012,RoyBondyopadhayaPRB2013,AlbrechtPRL2017,Liu_PRB2017}. Fig. \ref{SRK_dos}(a) shows the existence of local fermionic density at zero energy within the bulk gap. For $\phi_0=\pi/4$, DEC still shows quantised peak at zero-bias [see Fig. \ref{SRK_dIdV_V_0_025_01}(b)] and the subgap states remain very close to zero energy as one can see from Fig. \ref{SRK_dos}(b).  In Fig. \ref{SRK_dIdV_V_0_025_01}(b), we find that the height of ZBP peak corresponding to stronger coupling i.e. $\gamma'_p=0.25$ increases little bit as compared to the case of $\phi_0=0$, however, the height of ZPB peak corresponding to weak coupling $\gamma'_p=0.1$ almost remains same. We think this happens due to the combined effect of strong tunnel coupling and time-reversal breaking phase in the hopping term which can be interpreted as the effect of some effective local magnetic field because local magnetic field is known to increase the peak height as reported in Ref. \cite{RoyBondyopadhayaPRB2013}. These ZBP peaks corresponding to different tunnel strengths are not influenced equally by the TRS breaking phase due to the different levels of quasiparticle poisoning through non-identical tunnel junctions.

As discussed in the Ref. \cite{Degottardi_PRB2013}, if the TRS breaking perturbation is small, Majorana subgap modes remain at zero energy for SRK chain. However, as the strength of the TRS breaking term increases the zero-energy subgap edge-modes gradually become massive and the ZBP of DEC deviates from the quantised value [see Fig. \ref{SRK_dIdV_V_0_025_01}(c)]. Fig. \ref{SRK_dIdV_V_0_025_01}(d) depicts the complete absence of ZBP at $\phi_0=\pi/2$. This happens because at $\phi_0=\pi/2$, bulk gap closes and MBS vanishes identically. Here, DEC shows some equidistance peaks at the positions of the local density of states, as shown in Fig. \ref{SRK_dos}(d). 

We see that the strong tunnel coupling broadens the ZBP and smoothens the peak profile. For $\gamma'_p=0.25$, no significant reduction of peak-height is observed in ZBP for TRS breaking phase, when $\phi_0<\pi/2$. At $\phi_0=\pi/2$, finite-voltage DEC peak height increases due to the strong tunnel coupling, however no ZBP is observed, irrespective of the strength of tunnel couplings. 

In the case of the LRK chain, the effect of long-range interaction on DEC becomes dominant. It transforms the zero-energy Majorana modes to massive Dirac modes \cite{Viyuela_PRB2016,VodolaPRL2014}. Therefore, no ZBP peak is observed for any value of $\phi_0$ (See Fig. \ref{LRK_dIdV_V_0_025_01}). Fig. \ref{LRK_dos} represents energy level density profiles of the LRK chain for different values of TRS breaking phases: $\phi_0=0, \pi/4, \pi/3, \pi/2$. As depicted in Fig. \ref{LRK_dos}(a), massive subgap edge-states only exist in unbroken TRS case, i.e. $\phi_0=0$. Therefore, we observe unquantised peaks within the bulk gap. However, no subgap state is observed for all other values of $\phi_0$ (See Fig. \ref{LRK_dos}(b-d)). Hence, DEC does not show any peak within the bulk gap as Fig. \ref{LRK_dIdV_V_0_025_01}(b-d) represent. 
%\begin{figure}[htb]
%\centering{\includegraphics[width=1\linewidth]{Fourplots_LRK_beta_01.pdf}}
%%\centering{\includegraphics[width=1\textwidth]{Elec_cur_vs_V_epsi_0_gamma_05.pdf}}
%\caption{Plot of DEC ${dJ^e/dV}$ vs ${V}$ for LRK, ${{N}=40,|{\epsilon}|=0.0,|{\gamma_0}|=0.5,{\Delta}=0.15},\gamma_{p} =1.0 ,\gamma'_{p}=0.1, \alpha=0.5, \beta=0.5 $}
%\label{LRK_dIdV_V_0_01}
%\end{figure}

\begin{figure}[htb]
\centering{\includegraphics[width=1\linewidth]{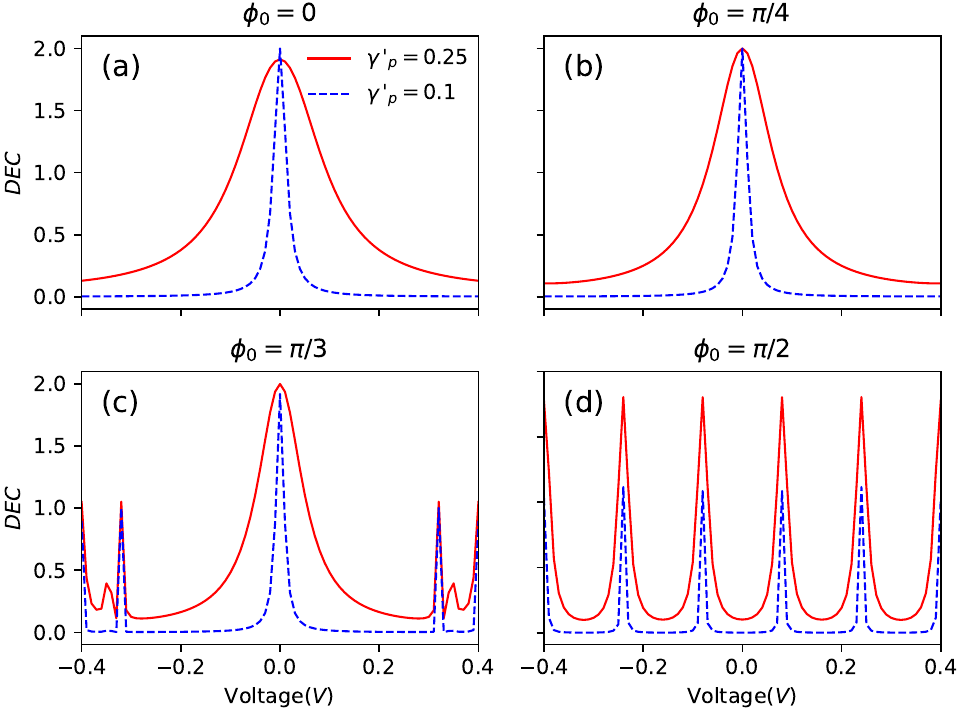}}
%\centering{\includegraphics[width=1\textwidth]{Elec_cur_vs_V_epsi_0_gamma_05.pdf}}
\caption{Plot of DEC (in units of $e^2/\hbar$) vs ${V}$ for SRK, ${{N}=40,|{\epsilon}|=0.0,|{\gamma_0}|=0.5,{\Delta}=0.15},, \alpha=10, \beta=10,\gamma_{p} =1.0. $}
\label{SRK_dIdV_V_0_025_01}
\end{figure}

\begin{figure}[htb]
\centering{\includegraphics[width=1\linewidth]{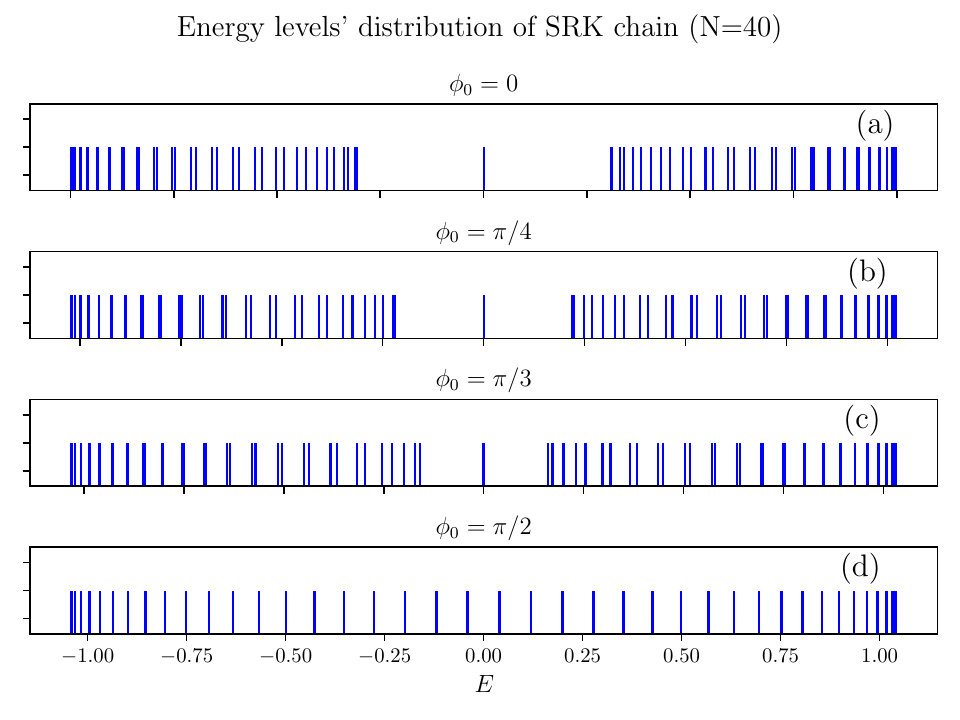}}
%\centering{\includegraphics[width=1\textwidth]{Elec_cur_vs_V_epsi_0_gamma_05.pdf}}
\caption{Plot of DOS for SRK, ${{N}=40,|{\epsilon}|=0.0,|{\gamma_0}|=0.5,{\Delta}=0.15}, \alpha=10, \beta=10,\gamma_{p} =1.0. $}
\label{SRK_dos}
\end{figure}

\begin{figure}[htb]
\centering{\includegraphics[width=1\linewidth]{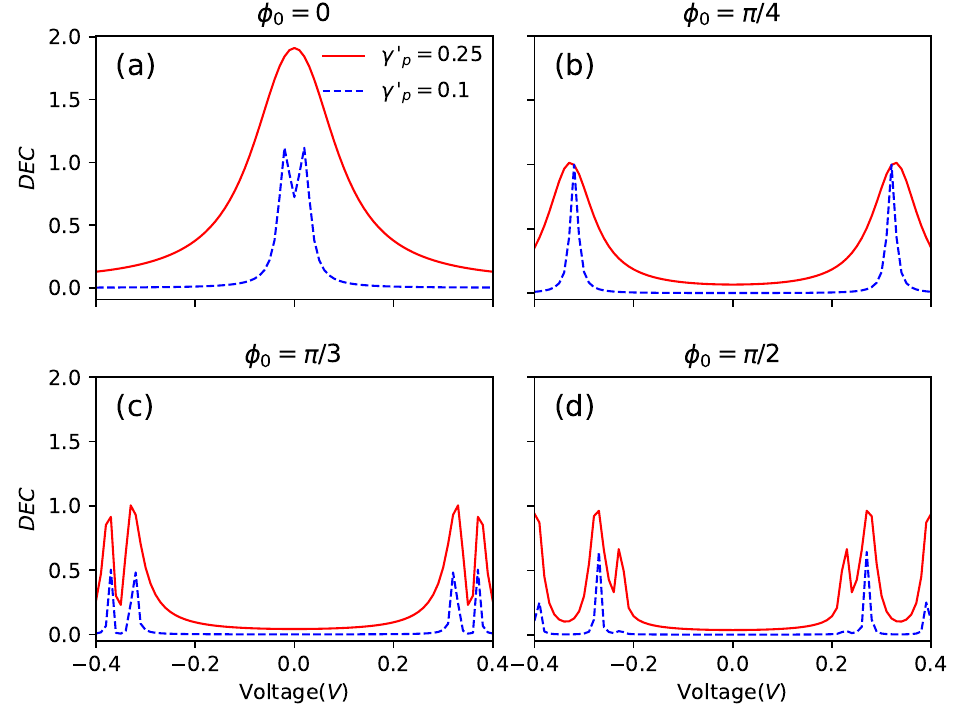}}
%\centering{\includegraphics[width=1\textwidth]{Elec_cur_vs_V_epsi_0_gamma_05.pdf}}
\caption{Plot of DEC (in units of $e^2/\hbar$) vs ${V}$ for LRK, ${{N}=40,|{\epsilon}|=0.0,|{\gamma_0}|=0.5,{\Delta}=0.15}, \alpha=0.5, \beta=0.5,\gamma_{p} =1.0. $}
\label{LRK_dIdV_V_0_025_01}
\end{figure}

\begin{figure}[htb]
\centering{\includegraphics[width=1\linewidth]{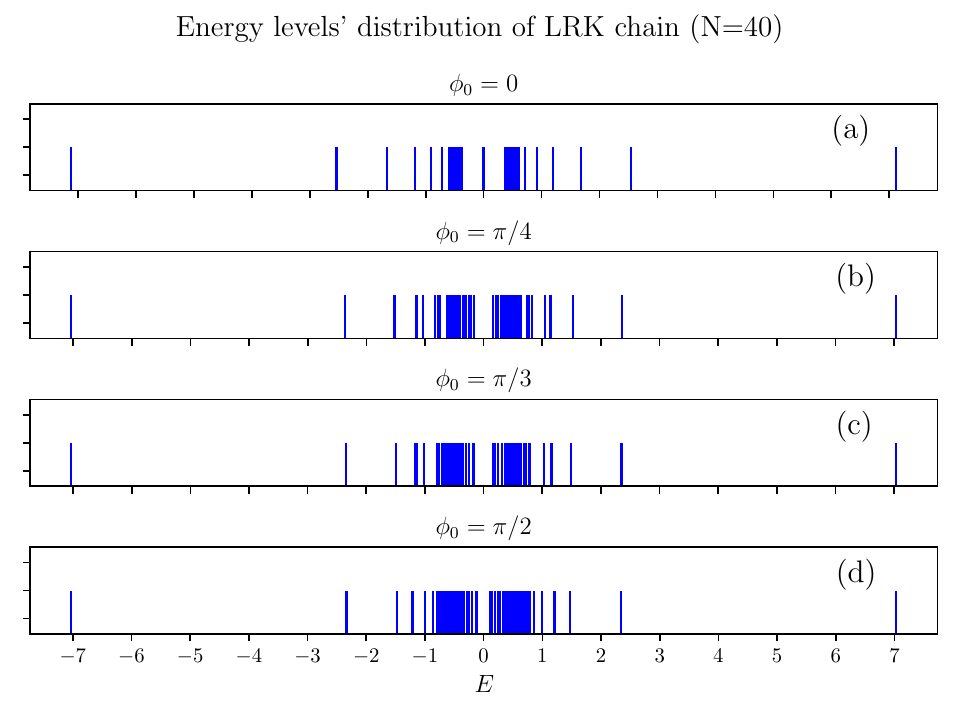}}
%\centering{\includegraphics[width=1\textwidth]{Elec_cur_vs_V_epsi_0_gamma_05.pdf}}
\caption{Plot of DOS for LRK, ${{N}=40,|{\epsilon}|=0.0,|{\gamma_0}|=0.5,{\Delta}=0.15}, \alpha=0.5, \beta=0.5,\gamma_{p} =1.0. $}
\label{LRK_dos}
\end{figure}

\end{subsection}

\begin{subsection}{Effect of TRS breaking on the localisation property of energy eigenstates}
\label{sec:TRS_IPR}
We find that TRS breaking affects the localisation property of the energy eigenstates for both SRK and LRK chains. To quantify the extent of delocalisation, we use Inverse Participation Ratio (IPR). IPR of $l$-th normalised eigenstate, $\psi_l$ is defined as \cite{Kramer_RPP_1993}
%\beq
%N_l^{\text{IPR}}= \frac{ \sum_{i=1}^{D} |c_i^{(l)}|^4 }{(\sum_{i=1}^{D} |c_i^{(l)}|^2)^2}\, ,
%\label{IPR}
%\eeq
\beq
N_l^{\text{IPR}}= \sum_{i=1}^{D} |c_i^{(l)}|^4 \, ,
\label{IPR}
\eeq
where, $\psi_l$ is expressed in a preferential basis $\{ |a_i \rangle \}_{i=1,\dots,D}$ as $\psi_l=\sum_{i=1}^{D} c_i^{(l)}|a_i \rangle $, and $D$ is the dimension of the Hamiltonian matrix. The higher (lower) the IPR value, the greater is the extent of localisation (delocalisation) of eigenstates. 
\begin{figure}[htb]
\vspace{1cm}
\centering{\includegraphics[width=1\linewidth]{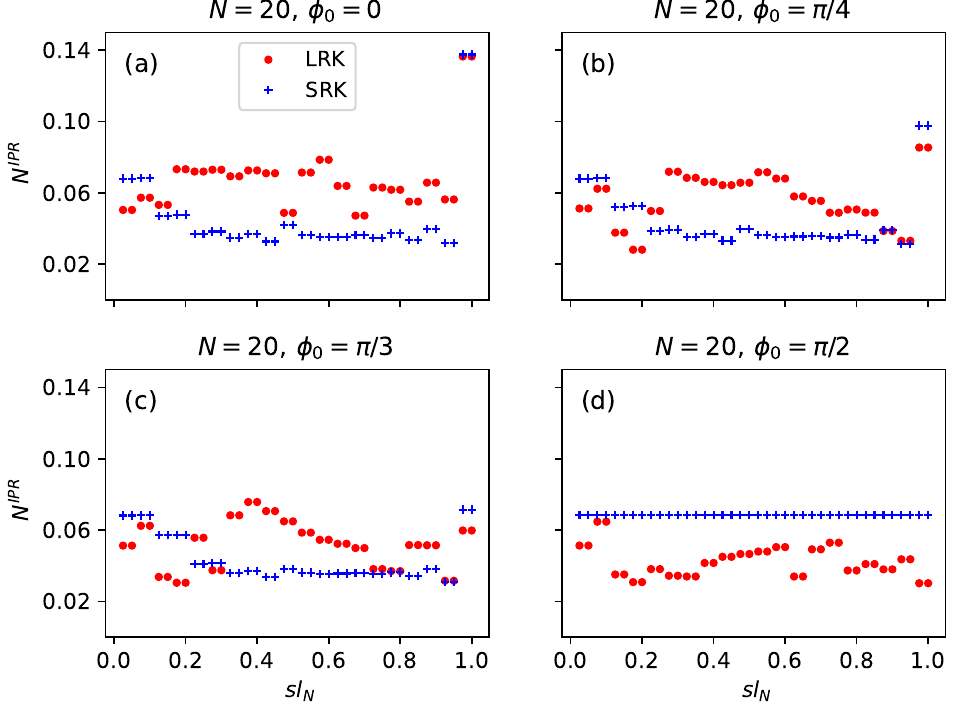}}
\caption{ IPR of different energy eigenstates of LRK and SRK chains with $N=20$. Here, normalised serial number ($sl_N$) is the serial
number of energy eigenstates divided by the total number of energy eigenstates. Two rightmost dots (pluses) corresponding to $sl_N=1.0~\text{and}~0.975$ represent subgap edge states for the LRK (SRK) chain in all the sub-figures.}
%\caption{Plot of electric current ${J^e}$ vs ${V}$ for SRK, ${{N}=20,|{\epsilon}|=0.0,|{\gamma_0}|=0.5,{\Delta}=0.15},\gamma_{p} =1.0 ,\gamma'_{p}=0.25, \alpha=10, \beta=10. $}
\label{IPR_20}
\end{figure}

\begin{figure}[htb]
%\vspace{1cm}
\centering{\includegraphics[width=1\linewidth]{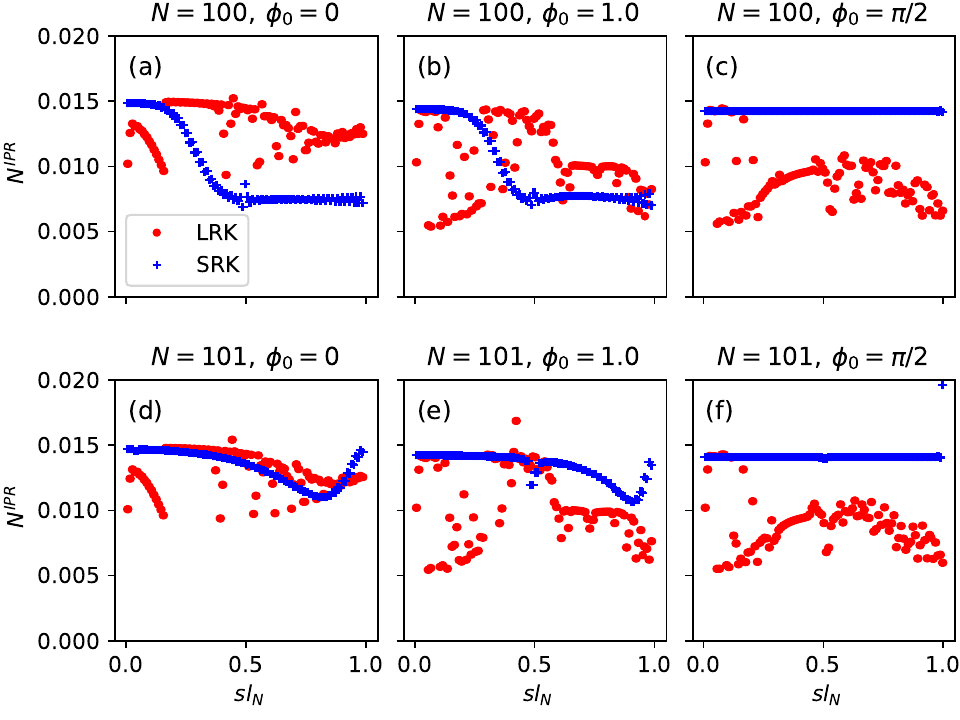}}
\caption{ IPR of bulk states of LRK and SRK chains for $N=100$ and $N=101$ respectively. $sl_N$ represents normalized serial number of the energy eigenstates}
%\caption{Plot of electric current ${J^e}$ vs ${V}$ for SRK, ${{N}=20,|{\e.psilon}|=0.0,|{\gamma_0}|=0.5,{\Delta}=0.15},\gamma_{p} =1.0 ,\gamma'_{p}=0.25, \alpha=10, \beta=10. $}
\label{IPR_100_101}
\end{figure}

Variations of IPR with TRS breaking phase ($\phi_0$) are depicted in Fig. \ref{IPR_20} and Fig. \ref{IPR_100_101}. In Fig. \ref{IPR_20}, we observe that the IPRs of the bulk states of the SRK chain do not change significantly as $\phi_0$ increases. However, the IPR of the subgap edge states (represented by two "+" signs at the top-right corner of Fig. \ref{IPR_20} (a)) decreases as $\phi_0$ increases ( see Fig. \ref{IPR_20} (b),(c))). These subgap edge states can be identified by the normalised serial numbers, $sl_N=1.0$ and $0.975$, respectively. Decrement of IPR value implies that the subgap edge states of the SRK chain which represent Majorana edge states for $\phi_0=0$, delocalise as the TRS breaking phase scales up however the bulk states of SRK do not show any significant change in the localisation property as $ \phi_0 \rightarrow \pi/2$. An interesting thing occurs to the SRK chain when $\phi_0 = \pi/2$. At this particular phase (see Fig. \ref{IPR_20}(d)), IPRs of different eigenstates (both bulk and edge modes) of the SRK Hamiltonian become almost equal which implies their equal delocalisation over the sites in the Hilbert space. It can be noted from the Fig. \ref{IPR_20}(d) that the $N^{IPR}$ values of eigenstates of the SRK chain are higher than the IPR values of the eigenstates of the SRK for $\phi_0 \neq \pi/2$ phases (see Fig. \ref{IPR_20}(a-c)). Therefore the eigenstates of SRK chain in $\phi_0=\pi/2$ phase are relatively localised compared to the eigenstates of SRK chain in other phases when $\phi_0 \neq \pi/2$. Further, we notice that for the SRK chain of any arbitrary length ($N$), the IPR values of all the eigenstates of that chain become almost same in the particular TRS breaking phase when $\phi_0=\pi/2$. This is an intriguing property of the SRK chain in the maximally TRS breaking phase i.e. $\phi_0=\pi/2$.

In the case of LRK chain, the overall IPR of the eigenstates decreases as the TRS breaking phase, $\phi_0$ grows from $0$ to $\pi/2$. It implies relatively stronger delocalisation of eigenstates as the TRS breaking phase increases and peaks at $\phi_0=\pi/2$. This is quite the opposite to the IPR pattern of the bulk states of the SRK chain. 

In Fig. \ref{IPR_100_101}, we represent the even-odd effect on the IPRs of bulk states. The upper row ( Fig. \ref{IPR_100_101}(a-c)) and lower row (Fig. \ref{IPR_100_101} (d-f)) corresponds to $N=100$ and $N=101$ sites chain, respectively. Comparing two rows of this figure, we observe that the IPRs distributions are different for the SRK chain with an even number of sites ($N=100$) and an odd number of sites ($N=101$) except for the maximally TRS breaking phase, $\phi_0=\pi/2$ in which IPR of the bulk states of SRK chain are almost identical. Comparing Fig. \ref{IPR_100_101} (a),(b) with Fig. \ref{IPR_100_101}(d),(e) we find that the bulk states of the SRK chain with an even number of sites (here, $N=100$) are more delocalised for all phases except $\phi_0=\pi/2$ as the average $N^{IPR}$ value is less for $N=100$ compared to $N=101$. However, in the case of $\phi_0=\pi/2$, no noticeable even-odd effect is observed in the delocalisation behaviour of the eigenstates as all the eigenstates acquire almost the same $N^{IPR}$ value (see Fig. \ref{IPR_100_101}(c),(f)). It is also observed that the bulk states of the SRK chain at maximally TRS breaking phase, $\phi_0=\pi/2$ are more localised than the bulk states in any other phases. This also corroborates our previous observation for $N=20$ case. 

However, for the LRK chain we do not observe any significant even-odd effect on the IPR of the eigenstates. This is quite expected because of the long-range interaction which makes fermion on each lattice site to interact with all other fermions placed at other sites. Therefore the collective interactions between all fermions distribute the eigenfunctions over the sites of the Hilbert space in a similar fashion for both $N=100$ and $N=101$ cases. Moreover, on contrary to the SRK chain, IPRs of different eigenstates of the LRK chain decrease as $\phi_0 \rightarrow \pi/2$ and the minimum value of IPRs is attained at $\phi_0 =\pi/2$. We can conclude that the eigenfunctions of the LRK chain become more delocalised as TRS breaking phase scales up from $0$ to $\pi/2$.

Variation of IPR with $\phi_0$ also affects junction electrical currents which we discuss in the next section.

\end{subsection}

\begin{subsection}{Electrical current}
\label{sec:EC}

Current vs voltage characteristics in an extended Kitaev chain show distinctive behaviours as TRS breaking phase, $\phi_0$ varies from $0$ to $\pi/2$. By varying the biasing voltage ($V$), we study the voltage dependence of junction electrical current ($J_e$) which is calculated by using Eq. \ref{elec_c}, for different values of $\phi_0$. As discussed before, LEGF formalism is employed here to calculate steady-state junction currents in open-system framework. Fig. \ref{SRK_I_V} and Fig. \ref{LRK_I_V} represent the current vs voltage characteristics for symmetrically biased SRK and LRK chains, respectively at zero on-site energy limit i.e. $\epsilon=0$. It can be noted that one needs to use the localisation property of the bulk states to understand the variation of magnitudes of saturation currents for SRK and LRK chains at different values of TRS breaking phase, $\phi_0$.

In Fig. \ref{SRK_I_V}, we plot $J_e$ for SRK chain which displays a sharp jump near the minuscule biasing voltage i.e. $V \in (0,0.1)$. The steepest jump is observed at $\phi_0=0$, and it gets less steep as $\phi_0 \rightarrow \pi/2$.  This happens because the subgap edge states become delocalised as $\phi_0$ increases which has  already been discussed in subsection \ref{sec:TRS_IPR} in connection with Fig. \ref{IPR_20}. Sharp rise of junction current which is directly related to the fermionic density of the subgap edge-states, falls as the delocalisation redistributes the edge-state fermionic density over a finite fraction of sites of the chain and converts the edge-states into extended states. 

As $V$ increases, $J^e$ shows a small plateau-like region for $V \in (0.1,0.3)$. This is the manifestation of gap in the energy spectrum (see Fig. \ref{SRK_dos}). This plateau like region is most prominent for $\phi_0=0$ which has the largest gap in the energy spectrum (see Fig. \ref{SRK_dos}(a)). This region is almost non-existent for $\phi_0=\pi/2$ (see Fig. \ref{SRK_dos}(d)), which does not show any plateau structure in $I$-$V$ characteristic. Moreover, $J^e$ displays saturation above $V=1.0$ and interestingly the saturation value of $J^e$ varies with the TRS breaking phase $\phi_0$. In SRK chain, the saturation current is maximum at $\phi_0=0$; it remains almost the same as $\phi_0$ increases, but it reaches a distinctly lower value at $\phi_0=\pi/2$. This saturation of current happens because no more energy level is available beyond the energy range $(-1.0\leq E \leq 1.0)$ as we see in Fig. \ref{SRK_dos}. However, to explain the variation of saturation current with $\phi_0$ in Fig. \ref{SRK_I_V}, we can refer to the IPR pattern of bulk states (see Fig. \ref{IPR_20}). Since the $N^{IPR}$ values of bulk states of SRK chain increase very slightly as $\phi_0 \rightarrow \pi/2$ ($\phi_0 \neq \pi/2$), a slight decrease is noticed in the saturation current. However, at the maximal TRS breaking phase ($\phi_0 = \pi/2$), $N^{IPR}$ of the bulk states of the SRK chain attain maximum values and thus become relatively localised (see Fig. \ref{IPR_20}(d)). These relatively localised bulk states of the $\phi_0=\pi/2$ phase hinder the transport process at finite voltage biasing and thus lower the value of saturation current which is represented by dotted curve in Fig. \ref{SRK_I_V}. 

Now, we analyse the Fig. \ref{LRK_I_V} which represents the $I$-$V$ characteristic for the LRK chain. Subgap edge states in the LRK chain become delocalised  rapidly as $\phi_0$ value increases. In Fig. \ref{IPR_20}, two isolated red dots corresponding to $sl_N=1.0 $ and $sl_N=0.975$ represent the LRK chain's subgap edge states. As $\phi_0$ increases, $N^{IPR}$ values corresponding to the edge states fall off which implies delocalisation. Therefore, the initial jump of $J^e$ in the $I$-$V$ characteristic is observed only in the TRS unbroken phase i.e. $\phi_0=0$ of the LRK chain. As we have seen before, a jump in $J^e$ at small voltage indicates presence of low-energy fermionic states at the edges which for $\phi_0=0$ are low-energy subgap edge states. As $V$ increases further, $J^e$ increases linearly until $V$ becomes sufficiently large to effectively envelope most of the energy levels which are depicted in Fig. \ref{LRK_dos} for $N=40$ case. At this point, $J^e$ shows saturation-like behaviour. Contrary to the SRK case, here we find the highest saturation current for $\phi_0=\pi/2$ and the lowest saturation current for $\phi_0=0$. This is just the opposite of the case of the SRK chain which we discussed in the last paragraph. Comparing Fig.  \ref{IPR_20}(a)-(d), one can understand that the overall $N^{IPR}$ 
%of the bulk states gradually decrease as $\phi_0$ increases from $0$ to $\pi/2$. Since, $N^{IPR}$ 
attains minimum values for $\phi_0=\pi/2$ in which the bulk states of the LRK chain are maximally delocalised. Therefore, this maximal TRS breaking phase of the LRK chain assists the electrical transport and causes the $J^e$ to saturate at maximum value.

\begin{figure}[htb]
\centering{\includegraphics[width=1\linewidth]{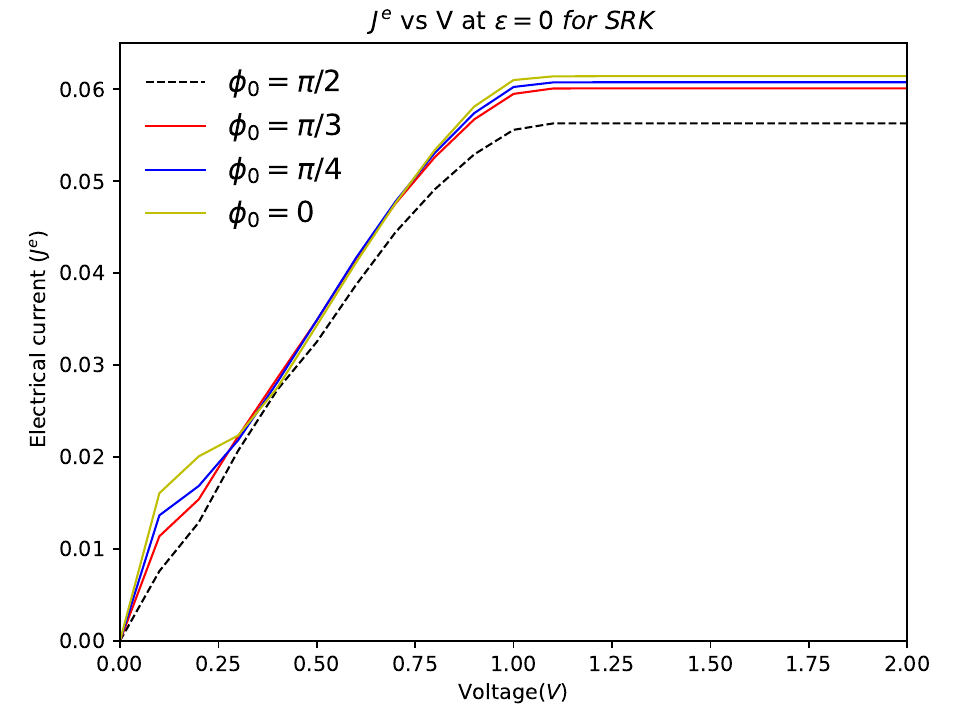}}
\caption{Plot of electrical current ${J^e}$ vs ${V}$ for SRK, ${{N}=20,|{\epsilon}|=0.0,|{\gamma_0}|=0.5,{\Delta}=0.15},\gamma_{p} =1.0 ,\gamma'_{p}=0.25, \alpha=10, \beta=10. $}
\label{SRK_I_V}
\end{figure}

\begin{figure}[htb]
\centering{\includegraphics[width=1\linewidth]{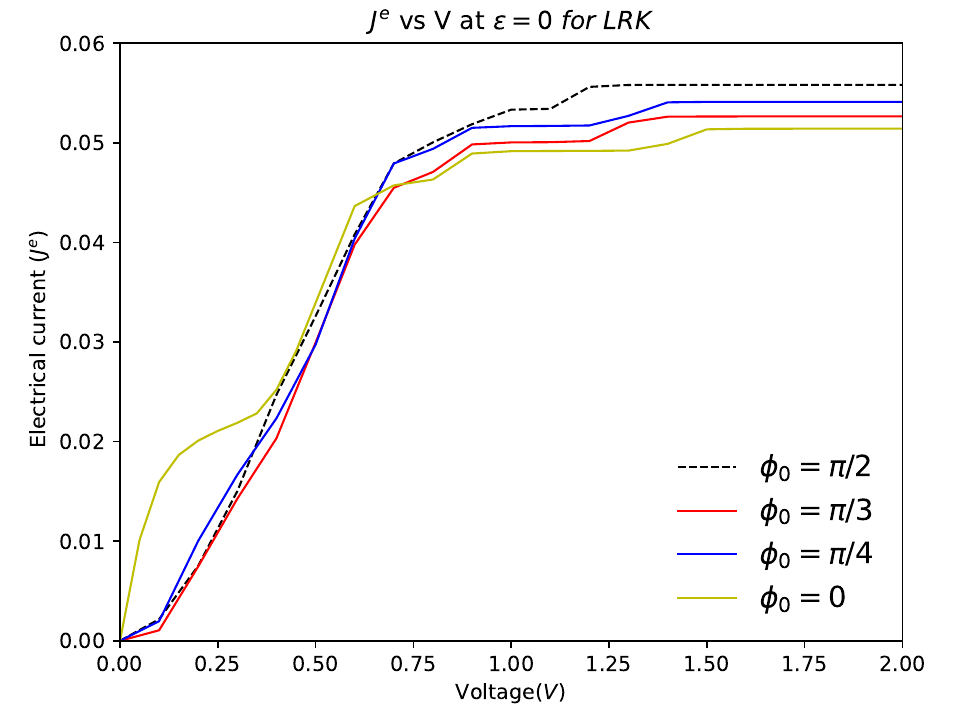}}
%\centering{\includegraphics[width=1\textwidth]{Elec_cur_vs_V_epsi_0_gamma_05.pdf}}
\caption{Plot of electrical current ${J^e}$ vs ${V}$ for LRK, ${{N}=20,|{\epsilon}|=0.0,|{\gamma_0}|=0.5,{\Delta}=0.15},\gamma_{p} =1.0 ,\gamma'_{p}=0.25, \alpha=0.5, \beta=0.5. $}
\label{LRK_I_V}
\end{figure}

%\begin{figure}[!htbp]
%    \centering
%    \includegraphics[width=6cm, height=5cm]{Electric_Current_0.0_TR.png}
%    \caption{Plot of electrical current ($J^e$) vs V ,
%N = 20; ${|\epsilon|}$  = 0.0}
%    \label{fig:1}
%    \centering
%    \includegraphics[width=6cm, height=5cm]{Electric_Current_1.0_TR.png}
%    \caption{Plot of electrical current ($J^e$) vs V ,
%N = 20; ${|\epsilon|}$  = 1.0}
%    \label{fig:2}
%\end{figure}
\end{subsection}

\begin{section}{Conclusion}
\label{sec:con}

We study differential electrical conductance and $I$-$V$ characteristics for varying strength of TRS breaking phase $\phi_0$. In addition, we also analyse the role of TRS breaking on transport for both LRK and SRK chains. In the case of the SRK chain, we found that the quantised zero-bias peak (ZBP) in DEC does not change significantly for $0 \leq \phi_0 < \pi/2$. This is consistent with the fact that the Majorana edge-states survive as long as the bulk gap does not close \cite{Degottardi_PRB2013}. However, at $\phi_0=\pi/2$, the bulk gap closes and the ZBP of DEC vanishes. While comparing with the LRK chain, we find that even the $\phi_0=0$ phase of the LRK chain lacks quantised ZBP due to the absence of zero-energy mode as a ramification long-range interaction. However, this phase may still show unquantised ZBP in the strong tunnel-coupling limit.

We observe the opposite behaviour of eigenstates of SRK and LRK chains with respect to their localization property as TRS breaking phase in the hopping term ($\phi_0$) gradually varies from $0$ to $\pi/2$. At the maximally broken TRS phase, i.e. $\phi_0=\pi/2$, bulk states of SRK chain show the lesser level of delocalisation, whereas that of the LRK chain shows a higher level of delocalisation. More interestingly, at this particular phase ($\phi_0=\pi/2$) all the eigenstates of the SRK chain show similar level of localisation as their IPR values become almost equal.

Difference in localisation property of eigenstates of LRK and SRK chains also lead to some changes in their respective $I$-$V$ characteristics. At the maximally TRS broken phase, i.e. $\phi_0=\pi/2$, this difference in $I$-$V$ becomes most prominent. In the case of SRK chain, $I$-$V$ characteristics in the low-voltage regime significantly differ at this phase. For the SRK chain at $\phi_0=\pi/2$, current saturates at the lowest value compared to TRS unbroken or other broken phases ($0 \leq \phi_0 < \pi/2$) due to the relatively localised eigenstates. On the other hand, for the LRK chain at $\phi_0=\pi/2$ which corresponds to higher delocalisation of eigenstates, the current saturates at a higher value compared to all other phases ($\phi_0 \neq \pi/2$). Therefore, the $I$-$V$ characteristic can be used to probe the contrasting behaviour of LRK and SRK chains at maximally TRS broken phase. We observe that the interplay between TRS breaking and long-range interaction reveals rich physics for extended Kitaev chain in TRS broken phase.

In the LRK model, the effect of TRS breaking is also very prominent in $I$-$V$ characteristics, especially at the low voltage region due to the absence of subgap states for any non-zero value of $\phi_0$. So, the initial sharp jump in current at low voltage which happens due to the presence of low-energy subgap states of LRK in the TRS unbroken phase, is entirely absent in the TRS broken phases. Thus, this $I$-$V$ characteristic can be regarded as an indication of TRS breaking in the LRK chain.

We find this study interesting because, firstly, we infer that a simple current-voltage characteristic which is a well-developed experimental technique, could be used to detect TRS breaking in both LRK and SRK chains. Secondly, we observe a novel phase of the short-range Kitaev chain at $\phi_0=\pi/2$, in which all the eigenstates undergo almost the same extent of delocalisation.

It may be noted that, apart from electrical current and differential conductance, other physical observables like thermoelectric current, thermal current and differential shot noise are gaining attention to understand properties of topological superconductors through quantum transport experiments \cite{Giuliano_PRB2018,SmirnovPRB2018,SmirnovPRB2019A,Smirnov_PRB2023,
BondyopadhayaJSP2022,Tsintzis_2022}. Therefore, it will also be interesting to study the effect of TRS on such physical observables.
\end{section}

\end{section}

\appendix
\setcounter{figure}{0}
\renewcommand\thefigure{A\arabic{figure}}
\section{Quantum Langevin equations and Green’s function method}
\label{App1}
It is well-known that the equilibrium transport in open-quantum systems can be studied by Quantum Langevin equations and Green’s function method (LEGF)\cite{DharSenPRB2006,DharJSP2006,RoyPRB2012,BondyopadhayaJSP2022,BhatDharPRB2020}.  In LEGF, we consider a N-TS-N type junction in which the middle superconducting chain is coupled to two metallic reservoirs on its two ends. In this paper, we follow the notation and the approach of Ref. \cite{DharSenPRB2006,BondyopadhayaJSP2022}. We consider the full system Hamiltonian as follows:
\bea
\mathcal{H}=\mathcal{H}_G+\mathcal{H}_M^L+\mathcal{H}_M^R+\Gamma_L+\Gamma_R,
\eea
where,
\bea
\mathcal{H}_G &=& \frac{1}{2}\sum_{l,m}\mathcal{H}_{lm}^{\rm G}a^{\dg}_la_m \, ,\nn \\
\mathcal{H}_{M}^L &=& \frac{1}{2}\sum_{\alpha,\beta}\, \mathcal{H}_{\alpha \beta}^{L}\, (a^L_\alpha)^{\dg}a^L_\beta \, , \nn  \\
\mathcal{H}_{M}^R &=& \frac{1}{2}\sum_{\alpha,\beta}\, \mathcal{H}_{\alpha \beta}^{R}\, (a^R_{\alpha})^{\dg}a^R_{\beta} \, , \nn \\
\Gamma_L &=& \frac{1}{2}\sum_{\alpha,l}\, \Gamma_{\alpha l}^{L}\, (a^L_{\alpha})^{\dg}a_{l} \, , \nn \\
\Gamma_R &=& \frac{1}{2}\sum_{m,\beta}\, \Gamma_{m \beta}^{R}\, a_{m}^{\dg}a^R_{\beta'} \, .
\label{GH}
\eea

These Hamiltonians are written in generalised basis ${\bf a}\equiv(a_1, a_2, \dots, a_{2N-1}, a_{2N})^T=(c_1, c^{\dg}_1, \dots, c_{N}, c^{\dg}_{N})^T$ for chain sites, and ${\bf a^p}\equiv(a_1^p, a_2^p, \dots)^T=(c_1^p, (c^p_1)^{\dg}, \dots)^T$ for reservoirs ($p=L,R$). In \ref{GH}, we use Greek indices ($\alpha, \beta$) for reservoir variables and Latin indices ($l,m$) for chain variables. Exact expressions for $\mathcal{H}_G$, $H_{M}^p$ and $\Gamma_p$ are given in Eq.\ref{HLRK}, Eq.\ref{HBath} and Eq.\ref{HTun}, respectively.

We assume that for times $t \leq t_0$ , the chain is disconnected from the reservoirs (each is in equilibrium at a specified temperature $T_p$ and chemical potential $\mu_p$ ). At time $t=t_0$, we connect both  reservoirs to the Kitaev chain through tunnel couplings to construct N-TS-N junction. Here, we are interested in the steady-state properties of this chain. For $t > t_0$, the Heisenberg equations of motion for the chain and reservoir variables read
\bea
\dot{a}_l &=& -i \sum_{m=1}^{2N} \mathcal{H}^{\rm G}_{lm} a_m -i\gamma_{L}' (a_2^L \, \delta_{l,2}-a_1^L \, \delta_{l,1}) \nn \\
&& -i\gamma_{R}'( a_2^R \, \delta_{l,2N}-a_1^R \, \delta_{l,2N-1}) \, ,
\label{eomwire}
\eea
for $l=1,\dots, 2N$, and 
\bea
 \dot{a}^L_{\alpha}  = -i \sum_{\beta} \mathcal{H}^{\rm L}_{\alpha \beta} \, a_{\beta}^{L} +i\gamma_{\rm L}'( a_1 \, \delta_{\alpha,1}-a_2 \, \delta_{\alpha,2}) \, , \nn \\
\label{eombathl}
 \eea
 for $\alpha = 1, \dots, \infty $, and
 \bea
 \dot{a}^R_{\alpha} = -i \sum_{\beta} \mathcal{H}^{\rm R}_{\alpha \beta }\, a_{\beta }^R
 +i\gamma_{\rm R}' ( a_{2N-1} \, \delta_{\alpha ,1}-a_{2N} \, \delta_{\alpha,2}  )\, , \nn \\
 \label{eombathr}
 \eea
for $\alpha =1 \dots, \infty$. We use retarded Green's function for isolated reservoir which are given below, to solve inhomogeneous equations  Eq.~\ref{eombathl} and Eq.~\ref{eombathr}. The single-particle retarded matrix Green's function of $p$-th reservoir is given by 
\beq
G^{p +}(t)=-i \theta(t) e^{-i \,\mathcal{H}^p t/\hbar} \, \nn 
\eeq
where, $G^{p +}$ and $\mathcal{H}^p$ are matrices and $\theta(t)$ is a step-function. Using these retarded Green's functions one can solve the equations of motion for reservoir variables ( Eq.~\ref{eombathl} and Eq.~\ref{eombathr}). Finally, substituting the solutions for reservoir variables into Eq.~\ref{eomwire}, one can rewrite Eq.~\ref{eomwire} in a form of generalised quantum Langevin equation in time domain which contains noise terms and self-energy terms \cite{RoyPRB2012,BondyopadhayaJSP2022}. These noise terms keep track of the non-equilibrium boundary conditions across the middle wire, which we impose in the beginning through the initial equilibrium correlators for reservoirs $\langle c^{p \dg}_\alpha (t_0)c^p_{\beta } (t_0) \rangle$ (\ref{corrl}) for $p=L, R$. On the other hand, the self-energy terms correspond to dissipation due to coupling with the reservoirs.

%Since, we are interested in the steady-state properties, we need to perform Fourier transformation of the aforesaid quantum Langevin equation. 
%
% which can
%be derived by integrating out the bath operators using
%the quantum LEGF formalism and considering the limit
%t 0 → −∞.

The quantum Langevin equations for chain variables can be solved in the frequency domain using the Fourier transformation. The Fourier transform of the chain variables are defined as $\tilde{a}_l (\omega)=\frac{1}{2\pi}\int_{-\infty}^{\infty} dt \, a_l(t)\, e^{i \omega t}$. Fourier transforms of reservoir Green's functions are  defined as
\beq
\tilde{G}^{p +}_{l,m}(\omega)=\int_{-\infty}^{\infty} dt e^{i \omega t} G^{p +}_{l,m} (t)\, . \nn
\eeq
Further, it can be shown that, within the bandwidth of the reservoir ($ |\omega| < 2 \gamma_p$), retarded Green's function for isolated reservoir are also given by \cite{RoyPRB2012},
\bea
\tilde{G}^{p +}_{1,1}(\omega) &=&\tilde{G}^{p +}_{2,2}(\omega)=\frac{1}{\gamma_p} \left[ \frac{\omega}{2 \gamma_p}-i\left( 1- \frac{\omega^2}{4 \gamma_p^2} \right)^{1/2} \right], \nn \\
\tilde{G}^{p +}_{1,2}(\omega)&=&\tilde{G}^{p +}_{2,1}(\omega)=0\, , \nn
\eea
where $p=L, R$.  Letting $t_0 \rightarrow -\infty$ and taking Fourier transform of the quantum Langevin equations for the chain variables, we get the following steady-state solutions for $\tilde{a}_l(\omega) $:
\begin{widetext}
\bea
\tilde{a}_l(\omega)&=&\sum_{m=1}^{2N} \tilde{G}^+_{l,m}(\omega)\left( \sum_{k=1,2} \tilde{\eta}_{k}^{\rm L} (\omega)\, \delta_{m,k} +\sum_{k=1,2} \tilde{\eta}_{k}^{\rm R} (\omega)\, \delta_{m,2N+k-2} \right) ,
\label{a1}
\eea 
\end{widetext}
where $l=1,\dots,2N $ and $\tilde{G}^+(\omega)$ is the retarded Green's function of the full system. Here, $\tilde{\eta}^{\rm L}_{k}(\omega) $ and $\tilde{\eta}^{\rm R}_{k}(\omega) $ ($k=1,2$) are the noise terms in the frequency domain, which are arising in the process of integrating out the variables of $L$ and $R$ bath, respectively. Noise-noise correlators in the frequency domain are given by \cite{DharSenPRB2006,BhatDharPRB2020,BondyopadhayaJSP2022}
\bea
\langle \tilde{\eta}^{\rm p \dg}_{1}(\omega)  \tilde{\eta}^{\rm p}_{1}(\omega ') \rangle &=& -(\gamma_p'^2/\pi) {\rm Im}[\tilde{G}^{p+}_{1,1}(\omega)] f(\omega, \mu_p,T_p) \delta(\omega- \omega') \, ,\nn \\
\langle \tilde{\eta}^{\rm p \dg}_{2}(\omega)  \tilde{\eta}^{\rm p}_{2}(\omega ') \rangle &=& -(\gamma_p'^2/\pi) {\rm Im}[\tilde{G}^{p+}_{2,2}(\omega)] f(\omega,-\mu_p,T_p) \delta(\omega- \omega')\, , \nn \\
\langle \tilde{\eta}^{\rm p \dg}_{1}(\omega)  \tilde{\eta}^{\rm p}_{2}(\omega ') \rangle & =& \langle \tilde{\eta}^{\rm p \dg}_{2}(\omega)  \tilde{\eta}^{\rm p}_{1}(\omega ') \rangle =0 \, . \nn \\
\label{nc}
\eea
Here, $\tilde{G}^{p +}_{l,m}(\omega)$ is the retarded Green's function of isolated bath ($p ={L,R}$). $\tilde{G}^{\rm L +}_{1,1}(\omega)$ and  $\tilde{G}^{\rm L +}_{2,2}(\omega)$
correspond to the first site of the left reservoir which is connected to the first site (left end) of the middle wire, whereas $\tilde{G}^{\rm R +}_{1,1}(\omega)$ and  $\tilde{G}^{\rm R +}_{2,2}(\omega)$ correspond to the first site of the right reservoir which is connected to the $N$-th site (right end) of the middle chain. 
%The detailed definitions and expressions of $\tilde{G}^{\alpha +}_{l,m}(\omega)$ are given in Appendix~\ref{App2}.

The retarded Green's function $\tilde{G}^+(\omega)$ of the full system in the frequency domain reads,
\bea
\tilde{G}^+(\omega)&=&{(\omega \one_{2N} - \mathcal{H}^G- \tilde{\Sigma}_{\rm L}^+(\omega) -\tilde{\Sigma}^+_{\rm R}(\omega))^{-1}} \nn \\
&=&{(\omega \one_{2N} - \tilde{\mathcal{H}})}^{-1}\,,
\label{FG0}
 \eea
where $\tilde{\Sigma}^+_{\rm p}$ is the self-energy correction to the Kitaev chain Hamiltonian originated from its interactions with the $p$-th reservoir. The effective Hamiltonian matrix of the Kitaev chain is given by $\tilde{\mathcal{H}}=\mathcal{H}^G+ \tilde{\Sigma}^+_{\rm L}(\omega) +\tilde{\Sigma}^+_{\rm R}(\omega)$.
The components of the self-energy terms $\tilde{\Sigma}^+_{\rm L/R}$ are as following:
\bea
[\tilde{\Sigma}_{\rm L}^+(\omega)]_{lm} &=& {\gamma_{\rm L}'}^2 \sum_{k=1}^{2} {\tilde{G}^{\rm  L +}_{k,k}(\omega)}\, \delta_{l,3-k}\, \delta_{l,m}\, ,\nn \\
{[\tilde{\Sigma}^+_{\rm R}(\omega)]}_{lm} &=& {\gamma_{\rm R}'}^2 \sum_{k=1}^{2}{\tilde{G}^{\rm R+}_{k,k}(\omega)}\,\delta_{l,2N-2+k}\, \delta_{l,m}\, , \nn \\
\eea
where $l,m= 1,\dots,2N $. Since $\tilde{\mathcal{H}}$ is a block diagonal matrix, numerical values of $\tilde{G}^+(\omega)$ can be calculated by inverting $(\omega \one_{2N} -\tilde{\mathcal{H}})$.

As we mentioned earlier, retarded Green's function method enables us to solve Eq. \ref{eombathl} and Eq. \ref{eombathl} for reservoir variables. Therefore, using Fourier transform we obtain the steady-state solutions of reservoir variables $\tilde{a}_l^{L/R}(\omega)$ defined at the edges of the reservoirs. For example, $\tilde{a}_{1}^L(\omega )$ and $\tilde{a}_{2}^L(\omega )$ for the left reservoir read 
\bea
\tilde{a}_{1}^L(\omega )\gamma_{\rm L}' &=&  -\tilde{\eta}^{\rm L}_1(\omega)-
\sum_{m=1}^{2}{[\tilde{\Sigma}^+_{\rm L}(\omega)]}_{1,m} \, \tilde{a}_{m}(\omega),\nn \\
 \tilde{a}_{2}^L(\omega )\gamma_{\rm L}' &=&  \tilde{\eta}^{\rm L}_2(\omega)+
\sum_{m=1}^{2}{[\tilde{\Sigma}^+_{\rm L}(\omega)]}_{2,m} \, \tilde{a}_{m}(\omega) \, .
\label{bathvariable}
\eea
These reservoir variables defined on the edges of the left and right reservoirs are essential to calculate junctions currents i.e. $J_{L/R}^e$  in a N-TS-N device. In order to calculate the steady-state junction currents, first we take the Fourier transformation of the currents expressions Eq.~\ref{elec_c}. Then substituting chain variables ($\tilde{a}_l(\omega)$) and reservoir variables ($\tilde{a}_\alpha^{L/R}(\omega)$) in the Fourier transformed expressions, and finally using noise-noise correlations for reservoirs (\ref{nc}), one can calculate the steady-state electrical currents ($J^e_{L/R}$) in N-TS-N junction \cite{BondyopadhayaJSP2022}. For example, the steady-state current in the left junction can be expressed in the frequency domain as
\bea
J_{\rm L}^e &=&  -2 e \gamma'_{L} \text{Im} [\langle c_{1}^{\dg} (t)c_{1}^L (t) \rangle ]= -2 e \gamma'_{L} \text{Im} [\langle a_{1}^{\dg} (t)a_{1}^L (t) \rangle ]  \nn\\
&=& -2e \gamma'_{L} \text{Im} \left[ \int^{\infty}_{-\infty} d \omega \, \int^{\infty}_{-\infty} d \omega' \, e^{(\omega -\omega')t} \langle \tilde{a}_{1}^{\dg} (\omega) \tilde{a}_{1}^L (\omega) \rangle \right] \nn \\
\eea
Now, substituting the values of $\tilde{a}_{1}^L (\omega)$, $\tilde{a}_l(\omega)$ from Eq. \ref{bathvariable}, Eq. \ref{a1} and finally, using the noise-noise correlations (\ref{nc}), one can calculate the steady-state current in the left junction. 

\bibliography{LRK_TRB}

\end{document}